# Results from the Telescope Array Experiment

**Gordon B. Thomson**[1]

*Department of Physics and Astronomy*
*University of Utah*
*Salt Lake City, UT, USA*
E-mail: `thomson@physics.utah.edu`

The Telescope Array (TA) is the largest experiment in the northern hemisphere studying ultrahigh energy cosmic rays. TA is a hybrid experiment, which means it has two detector systems: a surface detector and a fluorescence detector. In this paper we report on results from TA on the spectrum, composition, and anisotropy of cosmic rays. The spectrum measured by the TA surface detector, cosmic ray composition measured with the TA fluorescence detectors operating in stereoscopic mode, and a search for correlations between the pointing directions of cosmic rays, seen by the surface detector, and AGN's are presented.

## 1. Introduction

The Telescope Array (TA) experiment has the aim of studying ultrahigh energy cosmic rays, and is located in Millard County, Utah, USA. TA is a hybrid experiment which consists of a surface detector (SD) of 507 scintillation counters deployed on a 1.2 km grid, plus three fluorescence detector (FD) stations that overlook the SD. The two detector systems have been collecting data since early 2008. In this paper we present three physics results from the Telescope Array. First is the spectrum of cosmic rays measured by the SD. Here we use a technique new to the analysis of SD data. In our spectrum the ankle appears, plus a suppression (of significance 3.5 standard deviations) at the expected energy of the GZK cutoff [1]. Our spectrum is consistent with that measured by the High Resolution Fly's Eye (HiRes) experiment [2].

The composition of cosmic rays is studied most reliably by measuring the mean value of the depth of shower maximum, called <Xmax>. Here we measure <Xmax> using data from two of the TA FD's operating in stereoscopic mode. The result indicates that the composition is light, most likely mostly protons. This result bears on the controversy between the HiRes result [3], and that of the Pierre Auger experiment [4].

The third result presented here is a search for correlations between TA SD events' pointing directions and the positions of Active Galactic Nuclei (AGN) from the Veron-Cetty and Veron catalog [6]. We use the exact selection criteria of the Auger experiment [5], but find no correlations above the random level.

---

[1]     Speaker, representing the Telescope Array Collaboration.

## 2. Cosmic Ray Spectrum Measurement

In this section we describe a measurement of the spectrum of cosmic rays using the surface detector of the TA experiment. We use a technique new to SD analysis, where we make a detailed Monte Carlo simulation of our surface detectors which we compare to the actual data. We show that the simulation reproduces all the features of the data, and hence can be used to calculate the aperture of the surface detector. Previous experiments used only events whose energies are on the plateau of ~100% efficiency, but our analysis can include energies where the efficiency is much lower. In this way our measurement covers a wider energy range than that of many other experiments. We avoid the uncertainty in energy scale of shower simulation programs by normalizing our energy scale to the experimentally better controlled energy scale of our fluorescence detector using events seen in common. The result is a spectrum that agrees excellently with that of the HiRes experiment [2].

The problem with using shower Monte Carlo programs is that at ultrahigh energies generating showers requires too much CPU time. Many such programs include an approximation called "thinning" where, when particles fall below a preset energy, most are removed from further consideration and remaining particles at a similar point in phase space are assigned weights to take this into account. Using this approximation can lower the CPU requirements to a reasonable level. The resulting showers are accurate in the core region, but in the tails of the shower an inaccurate distribution of particles is made. We have developed a procedure called "dethinning" which attempts to replace the missing particles by changing a particle of weight $w$ into a swarm of $w$ particles; i.e., replacing the particles that were removed. We test this procedure by comparing the distributions of kinematic and dynamical variables from the shower with the TA SD data, and find excellent agreement. As an example, typical of many, we show in Figure 1 the zenith angle of cosmic rays. The data are the black points, and the Monte Carlo, normalized to the area of the data, is the red histogram. At this stage of our analysis we are considering cosmic rays of zenith angle less than 45°.

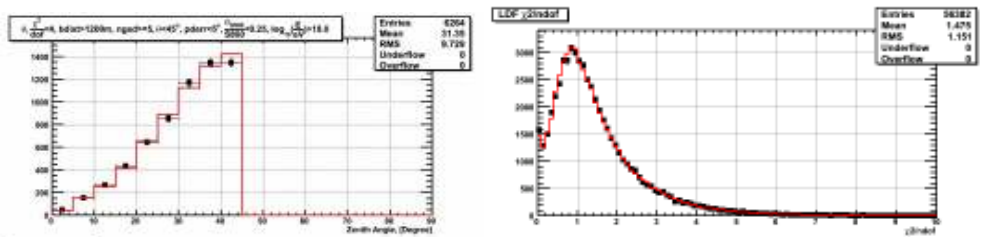

Figure 1. Comparison of TA surface detector data to the Monte Carlo simulation. The data is the black points and the Monte Carlo is the red histogram. In the left panel the zenith angle of cosmic rays is shown, and the right panel shows the reduced $\chi^2$ of the lateral distribution fit.

Since Monte Carlo programs use predictions of cross sections in the ultrahigh energy region made by using considerable extrapolation from actual measurements at lower energies, the energy scale of shower Monte Carlo programs has considerable uncertainty. Calculating that uncertainty is very difficult. To avoid this problem we adjusted the energy scale of our reconstruction of SD events to that of our fluorescence detector. Here one can reliably estimate the energy scale uncertainty, in our case about 20%. The adjustment consisted of a 27% reduction in event energies. We observe no nonlinearity in the comparison of the two energy scales.

Figure 2 shows the spectrum measured using TA surface detector data in black, and that of the HiRes experiment in color. The two spectra agree to a remarkable extent. This is particularly interesting because they were measured in different experiments using entirely different techniques. The TA SD spectrum shows the "ankle" feature at $10^{18.75}$ eV, and evidence for a flux suppression at $10^{19.75}$ eV, the energy expected for the GZK cutoff. The significance of the suppression is about 3.5 standard deviations.

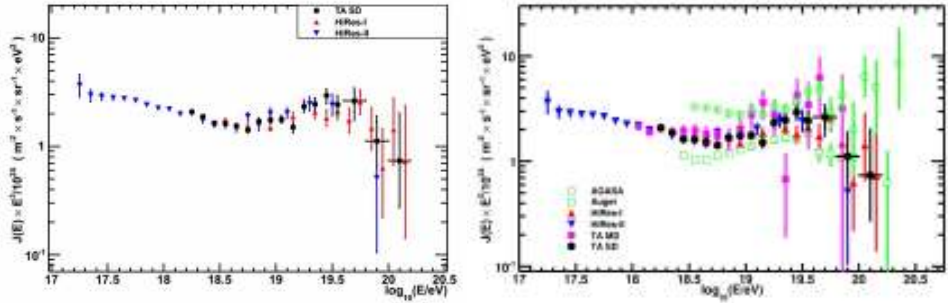

Figure 2. Left panel: the spectrum of cosmic rays measured using the Telescope array surface detector data, shown in black. The two HiRes monocular spectra are shown in red and blue. $E^3$ times the flux is shown. The agreement is remarkable, considering that the results come from different experiments which used very different techniques. For comparison, in the right panel results from the AGASA and Pierre Auger experiments are shown as green circles and green squares, respectively.

## 3. Cosmic Ray Composition Measurement

The most direct way of studying the composition of cosmic rays is to measure the mean depth of shower maximum, called <Xmax>. As an example of the sensitivity to composition, protons are expected to have <Xmax> 75-100 $g/cm^2$ deeper into the atmosphere than iron, where resolutions in <Xmax> of 20 $g/cm^2$ are achievable. We have measured <Xmax> using events seen by two of the TA fluorescence detectors operating in stereoscopic mode. This mode results in reconstruction of events' pointing directions with resolution better than 1°, the level needed for accurate <Xmax> measurement.

Our analysis proceeds in a way similar to that described in the section above, in that we made a detailed Monte Carlo simulation of the experiment (this time of the fluorescence detectors), and tested it using comparisons between data histograms of many kinematic and dynamic variables and Monte Carlo. These comparisons show that our simulation is very accurate. When we reconstruct Xmax for data events and plot means as a function of energy we get the points in Figure 3. The lines are the predictions from our Monte Carlo simulation using several hadronic generator programs, where the Monte Carlo events have been reconstructed using the same program as the data. The data indicate that the composition is very light, most likely almost all protons.

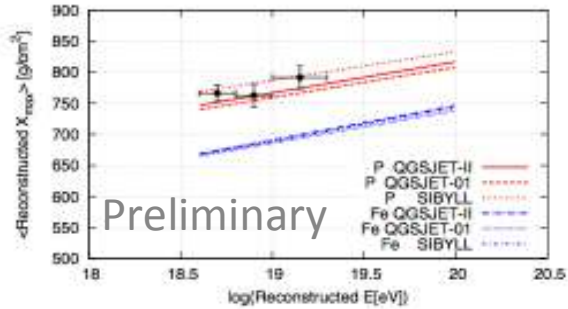

Figure 3. Mean Xmax as a function of energy. TA stereo data is shown, plus the predictions for protons and iron from several hadronic generator models.

## 4. Search for Correlations with Active Galactic Nuclei

The Pierre Auger collaboration has reported correlations between the pointing directions of their first 27 events above 57 EeV and active galactic nuclei (AGN) [5]. They scanned the first 14 events in energy, correlation angle, and maximum redshift to form a hypothesis of how to find the highest correlations, then tested this hypothesis with their next 13 events. In Reference [5] they state that correlations were found in this "test set" with a chance probability of 0.002.

Further data are available from the HiRes experiment [7]. Using the exact Auger selection criteria, they have 13 events above 57 EeV. 2 were correlated with AGN's with 3 expected by chance. This is consistent with no correlations.

Using exactly the same selection criteria and the same AGN catalogue [6] the TA collaboration searched for such correlations. We used the TA SD data set, described above, where events above 57 EeV have an angular resolution of about 1.2°, much smaller than the 3.1° correlation angles in the optimized Auger search criteria. Above 57 EeV we have 13 events, and observe 3 correlated events where 3.0 are expected by chance. There is thus no correlation signal in the TA data.